\documentclass[]{rspublic}

\usepackage{graphicx} 
\begin{document}

\newcommand{\ea}{{\it et al. }}
\newcommand{\apj}{{\it Astrophys. J.}}
\newcommand{\apjs}{{\it Astrophys. J. Suppl. Ser.}}
\newcommand{\aj}{{\it Astron. J.}}
\newcommand{\mnras}{{\it Mon. Not. R. Astron. Soc.}}
\newcommand{\aanda}{{\it Astron. Astrophys.}}
\newcommand{\ptra}{{\it Phil. Trans. R. Soc. A}}

%Some custom definitions:
\def\lsun{L$_\odot$}
\def\lsunsp{L$_\odot$ \ }
\def\msun{M$_\odot$}
\def\msunsp{M$_\odot$ \ }
\def\cc{cm$^{-3}$}
\def\kms{km\thinspace s$^{-1}$\ }
\def\gapprox{$_>\atop{^\sim}$}
\def\lapprox{$_<\atop{^\sim}$}
\def\co{$^{12}$CO\ }
\def\13co{$^{13}$CO}
\def\ceiosp{C$^{18}$O\ }
\def\ceio{C$^{18}$O}
\def\Av{$A_V$\ }

\title[The physics and modes of star cluster formation:
observations]{The physics and modes of star cluster formation:
observations}

\author[C.~J.~Lada]{Charles J. Lada}

\affiliation{Harvard--Smithsonian Center for Astrophysics, 60 Garden
Street, Cambridge MA 02138, USA}

\label{firstpage}

\maketitle

\begin{abstract}{\bf star formation; cluster formation} Stellar
clusters are born in cold and dusty molecular clouds and the youngest
clusters are embedded to various degrees in dusty dark molecular
material. Such embedded clusters can be considered protocluster
systems. The most deeply buried examples are so heavily obscured by
dust that they are only visible at infrared wavelengths. These
embedded protoclusters constitute the nearest laboratories for direct
astronomical investigation of the physical processes of cluster
formation and early evolution. I review the present state of empirical
knowledge concerning embedded cluster systems and discuss the
implications for understanding their formation and subsequent
evolution to produce bound stellar clusters.
\end{abstract}

\section{Introduction}

The question of the origin of stellar clusters is an old one. As early
as 1785, William Herschel first considered this problem in the pages
of these {\it Transactions} when he speculated on the origin of the
clusters Messier 80 and Messier 4 in Ophiuchus. More than two
centuries later, despite profound advances in astronomical science, we
find that the question of the physical origin of stellar clusters
remains largely a mystery. This is, in part, due to the fact that
cluster formation is a complex physical process that is intimately
linked to the process of star formation, for which there is as yet no
complete theory. The development of a theoretical understanding of
star and cluster formation therefore depends on first acquiring
detailed empirical knowledge of these phenomena. This can be a very
difficult task. Consider, for example, the globular clusters, the most
massive stellar clusters in our own Galaxy, the Milky Way. These
stellar systems are more than 12 billion years old and are no longer
formed in the Milky Way. Consequently, direct empirical study of their
formation process is not possible. The situation is considerably
better for Galactic open clusters. These typically span ages from 1
Myr to 1 Gyr and thus must be continually forming in the Milky Way,
making direct observational study of their formation process possible,
at least in principle. However, such studies have been seriously
hindered by the fact that open clusters are born in molecular clouds
and, during their formation and early evolution, are completely
embedded in molecular gas and dust. They are thus obscured from view
at optical wavelengths, where the traditional astronomical techniques
are most effective.

Fortunately, molecular clouds are considerably less opaque at infrared
wavelengths and the development and deployment of infrared imaging
cameras on optical- and infrared-optimized telescopes during the past
two decades has provided astronomers with the ability to detect,
survey and systematically study the extremely young embedded stellar
clusters within nearby molecular clouds. Such studies indicated that
embedded clusters are quite numerous and that they account for a
significant fraction, if not the majority, of all star formation
presently taking place in the Galaxy. Similarly, as discussed in
accompanying articles by de Grijs (2010) and Larsen (2010), young
clusters also appear to account for a significant fraction of star
formation in other galaxies, such as in the very active and luminous
starburst galaxies and in the vigorous star-formation episodes which
accompany galaxy mergers and close interactions. In our Galaxy, the
embedded-cluster phase lasts only 2--4 Myr and the vast majority of
embedded clusters which form in molecular clouds dissolve within 10
Myr or less of their birth. High cluster infant mortality has also
been inferred for clusters in other galaxies (see, e.g., de Grijs
2010; Larsen 2010). This early mortality of embedded clusters is
likely a result of the low star-formation efficiency that
characterizes the massive molecular-cloud cores within which the
clusters form. The physical reason for the low formation efficiencies
is not well understood and the origin of the massive cores themselves
is one of the many mysteries of modern astrophysical research.

The embedded clusters are the primary laboratory for research into the
question of the physical origin of stellar clusters. At the present
time, the answer to this question is shrouded by the dusty veils of
the dark molecular clouds in which stars and clusters are born. The
ultimate goal of a theoretical understanding of the physical process
of cluster formation is far from being realized. The first step
towards achieving this goal is to construct a solid empirical
foundation upon which a physical theory can eventually be built. This
then requires a detailed understanding of the physical nature of
embedded clusters and the environments in which they form. In this
article, I will review the empirical knowledge that is setting the
stage for the eventual development of a theory of cluster formation. I
will describe existing knowledge concerning many of the basic physical
properties of the embedded-cluster population, including the cluster
mass function, cluster birthrate, structural properties, etc., as well
as the nature of the gaseous cloud cores in which they form. I will
(at the end) briefly speculate on the implications of these results
for understanding the origin of stellar clusters. A more theoretical
discussion of possible physical mechanisms for forming such clusters
is contained in the accompanying article by Clarke (2010).

\begin{figure}
\vspace{-3.5cm}
\begin{center}
\includegraphics[scale=0.55, angle=90]{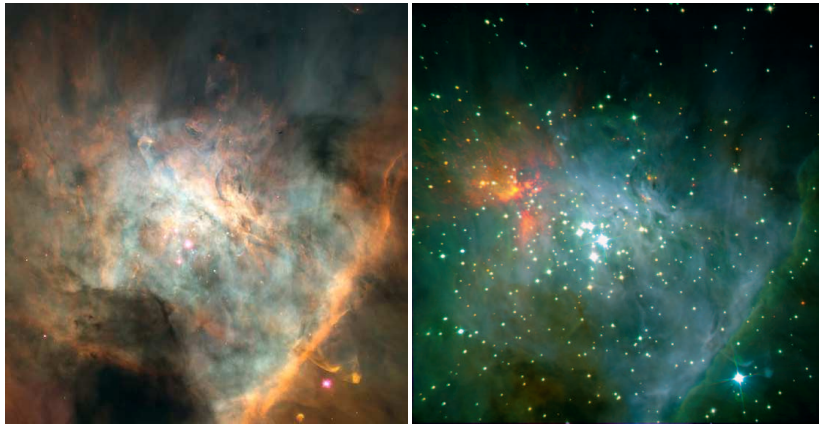}
\end{center}
\vskip -1.0in
\caption{{\it (left)} Optical and {\it (right)} infrared view of an
embedded Trapezium cluster associated with the Great Orion
Nebula. North is on the left, west at the top. (From Lada \& Lada
2003.)}
\end{figure}

\section{Demographics of embedded clusters}

\subsection{Definitions}

It is useful to begin any discussion of embedded clusters with the
definition of what constitutes such objects. Following Lada \& Lada
(2003), we first define a cluster to be a physically related group of
stars that satisfy the following two requirements. First, the group
must have a stellar-mass volume density sufficiently large to render
it stable against tidal disruption by the Galaxy (i.e., $\rho_* \geq
0.1$ \msunsp pc$^{-3}$; Bok 1934) and by passing interstellar clouds
(i.e., $\rho_* \geq 1.0$ \msunsp pc$^{-3}$; Spitzer 1958). A group of
only 8--10, 0.5 \msunsp stars within a radius of 1 pc would satisfy
this first requirement. Second, the group should contain a sufficient
number of members to ensure that, if in a state of virial equilibrium,
its evaporation time (i.e., the time it takes for internal,
gravitational stellar encounters to eject all of its members) is
greater than the typical lifetime of Galactic open clusters of $\sim
10^8$ yr. For a cluster in virial equilibrium, this number is about 35
stars. An embedded cluster is then defined as a cluster that is fully
or partially embedded in interstellar gas and dust. The above
definition of a cluster includes both gravitationally bound and
unbound stellar systems. Bound stellar clusters are those whose total
energy (kinetic and potential) is negative. When determining the total
energy of a cluster, we take into account contributions from all forms
of mass within the boundary of the cluster, including interstellar
material. This is crucial in evaluating the dynamical states of
embedded clusters, since generally the mass of interstellar material
exceeds the stellar mass within the cluster boundaries.

\subsection{Identification: infrared imaging surveys}

Because most of their members are often optically obscured by
interstellar dust, infrared imaging is generally required to identify
and investigate embedded clusters. The existence of a cluster is
established observationally by an excess density of stars over the
background. Consequently, the detectability of a cluster depends on
its richness and compactness, the brightness of the cluster members
and the density of background objects. Figure 1 shows optical and
infrared images of the famous Orion Nebula and its embedded cluster.
It clearly illustrates the advantage of infrared imaging for revealing
the stellar content of embedded clusters. A number of authors have
compiled catalogues of embedded clusters based on infrared
observations. Porras \ea (2003) compiled a list of 34 clusters, which
is likely nearly complete for clusters within 1 kpc of the sun. Lada
\& Lada (2003) list 76 embedded clusters with well-determined
properties within 2.5 kpc of the sun. They estimate that their list is
representative but incomplete and that there are likely a total of
about 200 such clusters within this volume of space, most of which are
unobserved to date. Bica \ea (2003) compiled a highly incomplete list
containing more than 300 infrared clusters, stellar groups and
candidate clusters, most within about 5 kpc of the sun, some fraction
of which are likely embedded. Even cursory consideration of these
studies suggests that embedded clusters are quite numerous in the
Galactic disc.

\subsection{Membership: signatures of youth from infrared to X-rays}

Although the existence of a cluster can be established by the increase
of source density over the background, identification of the
individual members is considerably more difficult, especially for
intrinsically faint members whose numbers are only comparable or
significantly lower than those of the background/foreground stellar
field population. The size of the cluster's membership can be
determined statistically through comparison with the
background/foreground field-star surface density over the cluster
area. However, determining whether or not a specific star is a cluster
member requires independent information. Since embedded clusters are
very young stellar systems, independent indicators of stellar youth,
such as the presence of circumstellar discs, variable emission lines,
X-ray emission, etc., can be employed to ascertain membership of
individual stars.

For typical embedded-cluster ages of 1--3 Myr, at least half of the
members will be surrounded by circumstellar discs, remnants of the
star-formation process (e.g., Haisch \ea 2001; Hernandez \ea
2008). Many of these discs will be accretion discs and exhibit strong,
variable hydrogen recombination lines at optical and infrared
wavelengths (e.g., Green \& Lada 1996; White \ea 2007). Infrared
spectroscopic surveys, especially those done with multi-object
spectrographs, can be used to identify accreting objects in a cluster
field. However, such observations can be time consuming and even
prohibitive for faint members, which can account for a significant
portion of the cluster population.

\begin{figure}
%\hskip 0.83in
\begin{center}
\includegraphics[scale=0.35]{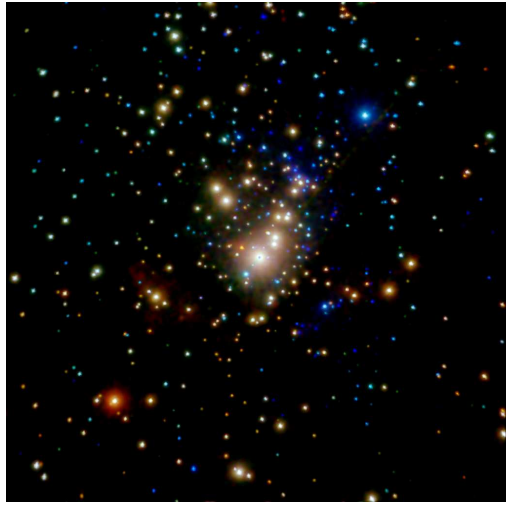} 
\end{center}
\caption{X-ray image of the Trapezium cluster in Orion, obtained with
a deep integration using the {\sl Chandra} space
observatory. Essentially all cluster members were detected as X-ray
sources (Getman \ea 2005). North is at the top, west on the right.}
\end{figure}

Circumstellar discs can be detected directly with infrared broad-band
photometry as a readily measurable excess above the expected emission
from a normal stellar atmosphere. Excesses due to circumstellar discs
and envelopes are most robustly measured at longer infrared
wavelengths (i.e., $\lambda \geq 2 \mu$m). In recent years, the {\sl
Spitzer Space Telescope}, operating at wavelengths between 3 and 70
$\mu$m, has provided the capability to survey and measure infrared
excesses and circumstellar-disc emission with unprecedented
sensitivity across the complete mass range of stars in
embedded-cluster regions (e.g., Allen \ea 2007). These observations
are providing the first statistically complete accountings of disc
populations in individual young clusters (e.g., Lada \ea 2006;
Sicilia--Aguilar \ea 2006) and important new insights into their
structural properties (Gutermuth \ea 2009).

Although infrared observations are sufficiently sensitive to detect
the presence of infrared excess from stars across the entire spectrum
of stellar mass, there are typically many cluster members which have
no discs or detectable infrared excess, presumably because these stars
have passed quickly through the disc phase of early stellar
evolution. As many as 50\% of all members of a young cluster may have
lost their discs and any other accretion signatures at an age of 2--3
Myr. Observations made from space with X-ray observatories such as the
{\sl Chandra X-ray Observatory} have shown that young stars emit
X-rays at levels $10^2 - 10^4$ times that of normal stars,
particularly during the first 10 Myr of their lives (Feigelson \&
Prebisch 2005). Such emission appears to be detectable across almost
the entire stellar-mass spectrum, whether or not the stars have discs
(e.g., Telleschi \ea 2007). In principle, X-ray surveys can be used to
identify the discless membership of a young stellar population and
thus, in combination with infrared surveys, lead to identification of
all members of an embedded cluster. For example, a deep {\sl Chandra
Observatory} observation of the Trapezium cluster in the Orion Nebula
resulted in the X-ray image shown in figure 2, in which all cluster
members (with and without discs) are detected as X-ray sources (Getman
\ea 2005). However, these {\sl Chandra} observations were very deep,
and in general the degree to which X-ray surveys can sample the
membership of a cluster depends on the sensitivity and depth of the
observations. In typical situations, infrared observations are more
sensitive to the faintest members of a cluster population (e.g.,
Winston \ea 2007). Nonetheless, to accurately obtain a census of the
membership of an embedded cluster requires a varied set of
observations and such information is, at the present time, only
available for a limited set of clusters.

\section{Fundamental physical properties of embedded clusters}

\subsection{Embedded cluster mass spectrum}

The frequency distribution of embedded-cluster masses is a fundamental
property of these systems which any physical theory of cluster
formation must explain. Lada \& Lada (2003) used the data from their
cluster catalogue to uniformly derive masses for the known embedded
clusters within 2.5 kpc. They then constructed the embedded-cluster
mass spectrum (ECMS) and found it to be given by a power law of the
form $\mathrm{d}N/\mathrm{d}M \propto M^{-\alpha}$, where $\alpha
\approx 1.7$--2.0 for masses in excess of about 50 \msun. For a
distribution with $\alpha = 2$, the total mass of clusters in any
logarithmic interval (bin) is constant. Thus, even though there are
many more 50 \msunsp than 1000 \msunsp clusters, the very rare 1000
\msunsp clusters contribute the same fraction of the total mass in
clusters as the more numerous 50 \msunsp clusters. This, in turn,
implies that if a star in the Galaxy is born in a cluster, it is just
as likely to form in a 1000 \msunsp as in a 50 \msunsp cluster. The
ECMS very likely represents the primordial mass function of Galactic
open clusters. In this context, it is interesting to note that a
recent attempt to determine the mass function of classical open
clusters also produced a power-law relation with an index of $\alpha
\approx 2$ (Piskunov \ea 2008; see also discussion by Larsen 2010).

Observations also suggest that the ECMS peaks near 50 \msun and then
falls off towards lower masses. Indeed, in a detailed study of all
known stellar groups and clusters within 1 kpc of the sun, Porras \ea
(2003) found that although small stellar ($N_*<$ 30) groups greatly
outnumber clusters, clusters with 100 or more members account for 80\%
of the total number of stars contained in both groups and
clusters. This is consistent with the notion that most stars that form
in the Galaxy form or have formed in embedded clusters with 100 or
more members, and these are the clusters with masses of 50 \msunsp or
greater.

\begin{figure}
%\hskip 0.35in
\begin{center}
\includegraphics[scale=0.45]{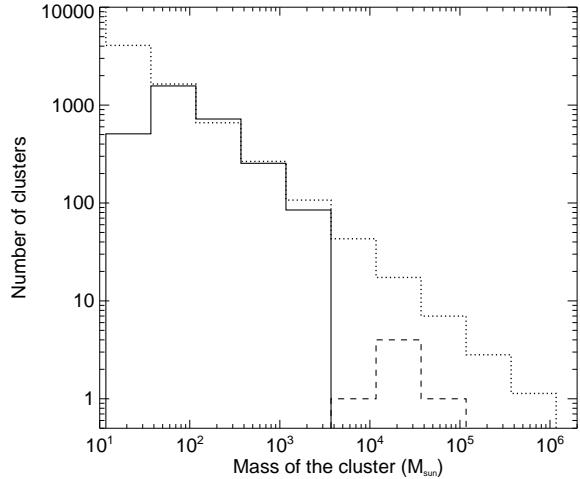}
\end{center}
%\hskip 1.8in
\caption{Embedded-cluster mass spectrum for the Galaxy. The scaled
ECMS for clusters from Lada \& Lada's (2003) compilation of cluster
masses within 2.5 kpc is plotted as the solid trace. The predicted
ECMS for all masses and a spectral index of $\alpha = -1.7$ is shown
as the dotted histogram. Massive embedded clusters from the list of
Ascenso (2008) are represented by the dashed trace. (Adapted from
Ascenso 2008.)}
\end{figure}

The most massive cluster in the Lada \& Lada (2003) sample was the
Orion Nebula Cluster, weighing in at a little above $10^3$ \msun. More
massive systems are known to exist, so it seems natural to assume that
the lack of massive clusters in the Lada \& Lada mass function is not
caused by a truncation at high masses but is instead a result of the
distance restriction of 2.5 kpc coupled with small-number statistics
at higher cluster mass. Recently, Ascenso (2009) compiled a list of
known clusters independent of distance in the Galaxy with masses
estimated as in excess of 10$^4$ \msun. The largest of these appears
to be Westerlund 1, whose mass of about $5 \times 10^4$ \msunsp
approaches that of globular clusters (Brandner \ea 2008). Of the nine
clusters in her list, six have ages of 3 Myr or less and can be
considered as part of the embedded, young cluster population. To
estimate the numbers of massive, embedded clusters that may inhabit
the Milky Way, Ascenso (2008) scaled the 2.5 kpc ECMS to correspond to
that expected for the entire area of the Galactic disc ($R_\mathrm{MW}
= 13$ kpc). Figure 3 shows the scaled ECMS, along with the six massive
clusters from Ascenso's list. The predicted distribution of clusters
is also plotted, calculated for a mass spectrum index $\alpha = 1.7$
and normalized to the scaled observed distribution at cluster masses
$\leq 10^3$ \msun (Ascenso 2008). A significant number of young
clusters are predicted for cluster masses in excess of $2-3 \times
10^3$ \msunsp, but only a few such clusters have been identified.

Understanding the nature and origin of high-mass ($M > 10^4$ \msun)
clusters is of considerable interest because observations indicate
that such large clusters are a major component of the vigorous
star-formation activity in merging and starburst galaxies (e.g.,
Bastian \ea 2005; see also accompanying reviews by de Grijs 2010;
Larsen 2010). Are there examples of such clusters being produced in
the Milky Way at the present epoch or is the ECMS truncated at high
mass? The question of whether there exists a high-mass cutoff or
truncation to the ECMS is difficult to answer. There are no known
embedded clusters with reliably determined masses in excess of $10^5$
\msun. One would think that such clusters would be easily identified
if they existed. However, the predicted numbers are low enough that
small-number statistics could be responsible, in part, for the
observed lack of such systems in the Galaxy today. The present-day
masses of classical open clusters offer additional constraints on the
possibility of a truncation of the upper end of the primordial
ECMS. For example, consider the most massive open clusters in our
Galaxy with ages $\geq 20$ Myr. Those systems all have masses $< 10^4$
\msunsp (e.g., Bruch \& Sanders 1983; Mermilliod 2000). Recently,
astronomers identified three red-supergiant-rich clusters with ages
between 10 and 20 Myr and masses between 2 and $4 \times 10^4$ \msun
(Figer \ea 2006; Davies \ea 2007; Clark \ea 2009). If the
cluster-formation rate were constant over time, figure 4 suggests that
we might expect at least 100 objects with masses $\geq 10^4$ \msunsp
and with ages of 300 Myr or less. So far, only three such objects are
known. Therefore, either the ECMS is truncated at masses $> 10^4$
\msunsp or such clusters did not survive emergence from molecular
clouds and dissolved very rapidly into the field.

\subsection{Ages, birthrates and infant mortality}

The theory of stellar structure and evolution predicts fairly
precisely the mass-dependent evolution of stellar luminosities and
temperatures with age. This enables the age of a stellar population to
be determined by placing the stars on a Hertzsprung--Russell (HR)
diagram, which plots the distribution of stellar luminosities versus
temperatures. The locations of stars are then compared to the
predictions of stellar-evolution theory. This has proved to be a
powerful tool for age dating mature clusters whose more massive stars
have finished their main hydrogen-burning stage and experience
significant luminosity and temperature evolution. However, age dating
young, embedded clusters has proved to be considerably more
difficult. The reason for this is twofold. First, the majority of
stars in an embedded cluster are pre-main-sequence (PMS) stars, that
is, stars that have not yet evolved to the point where nuclear burning
commences in their interiors. For these stars, the theoretical
trajectories on the HR diagram can be highly uncertain, especially for
stars with ages of 1 Myr or less. Second, it is difficult to measure
accurate luminosities and temperatures for PMS stars since they are
obscured by interstellar dust, exhibit infrared excess emission and
are variable. Nonetheless, a number of embedded clusters have been age
dated in this fashion and typical median ages of stellar members in
clusters range between 1 and 3 Myr, with the age spread in a given
cluster of comparable magnitude to its age (Lada \& Lada 2003). These
results are consistent with the early work of Leisawitz \ea (1989),
who found that clusters older than about 5 Myr exhibit no evidence of
interstellar molecular gas within or near their boundaries. Thus, the
embedded-cluster phase of evolution lasts somewhere between 3 and 5
Myr. [It is interesting to note here that recent analysis using only
main-sequence cluster members on the HR diagram by Naylor (2009)
yields age estimates 1.5 to 2 times older than derived from the PMS
stars.]

\begin{figure}
\begin{center}
\includegraphics[scale=.60]{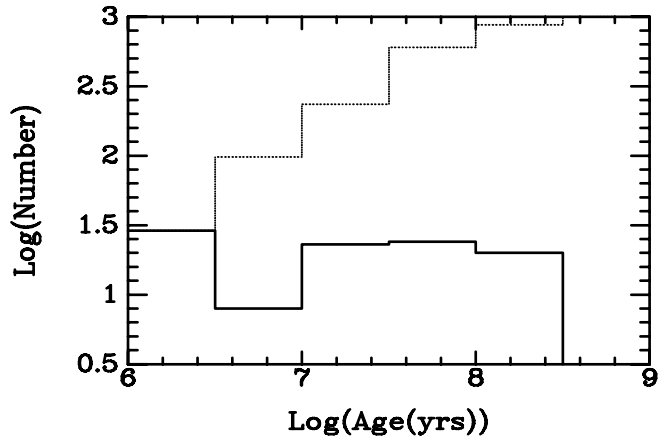}
\end{center}
\caption{Observed frequency distribution of ages for classical open
and embedded clusters within 2 kpc of the sun (solid line) compared to
that predicted for a constant rate of cluster formation (dotted
line). The discrepancy between the predicted and observed values in
the second bin indicates that the vast majority of embedded clusters
do not survive emergence from a molecular cloud. (From Lada \& Lada
2003.)}
\end{figure} 

Assuming typical ages of embedded clusters of between 1 and 2 Myr,
Lada \& Lada (2003) estimated their birthrate within 2 kpc from the
sun at $\geq 2$ to 4 Myr$^{-1}$ kpc$^{-2}$. Upon emergence from their
dusty wombs, embedded clusters become optically visible open
clusters. Yet, the embedded-cluster birthrate is an order of magnitude
higher than that (0.45 Myr$^{-1}$ kpc$^{-2}$), estimated for classical
open clusters within 2 kpc of the sun (Battinelli \& Capuzzo--Dolcetta
1991). The large difference in the estimated birthrates is key in
understanding the formation and early evolution of Galactic clusters.

The distribution of ages for a combined sample of embedded and open
clusters within 2 kpc is shown in figure 4. Embedded clusters fall
into the first bin and classical open clusters have ages that span a
range from roughly $10^6$ to $10^8$ yr. The distribution is
surprisingly flat. Also plotted is the predicted age distribution for
a constant rate of cluster formation, assuming all clusters can live
to ages of at least a few times $10^8$ yr. There is a large and
increasing discrepancy between the expected and observed numbers. This
figure suggests a high infant-mortality rate for embedded clusters.
Most of these clusters do not survive emergence from their parent
molecular cloud. Roughly 90\% dissolve in less than 10 Myr. Less than
about 4\% of the initial cluster population reach an age of 100 Myr.
Lada \& Lada (2003) suggest that only the most massive embedded
clusters ($M > 500$ \msun) survive to ages of 100 Myr or more. It has
been understood for some time that passing massive interstellar clouds
could tidally disrupt open clusters with mass densities $\leq 1$ \msun
pc$^{-3}$ within 200 Myr (Spitzer 1958; Gieles \ea 2006). These tidal
interactions with giant molecular clouds (GMCs) likely account for the
flat age distribution of clusters with ages $> 10$ Myr. However, the
rapid disintegration of most clusters in the first 10 Myr of life
requires another mechanism. To understand this process, we need to
first understand the conditions under which embedded clusters form.

\subsection{Association with molecular clouds}

The intimate physical association with interstellar gas and dust is
the defining characteristic of embedded clusters. Such objects are
immersed in dusty, gaseous material to varying degrees. The most
deeply embedded clusters are believed to be the youngest and least
evolved objects of this class. They are found within massive, dense
structures or cores inside GMCs. GMCs are, along with globular
clusters, the most massive objects in the Milky Way. They are complex
structures, often defined by long filamentary features. Such clouds
are primarily composed of molecular hydrogen gas and characterized by
a wide range of gas densities ($10^2 \leq n(\mathrm{H}_2) \leq 10^6$
cm$^{-3}$). Star formation has long been known to take place in the
dense ($n(\mathrm{H}_2) > 10^4 \mathrm{cm}^{-3}$) component of
molecular clouds (Lada 1992) and this component only accounts for
between 1 and 10\% of the total mass of GMCs observed in the vicinity
of the sun. There are also indications that the star-formation rate in
local clouds is closely related to their total content of dense gas
(Lada \ea 2009). The latter is found in the form of filaments and also
in more compact cores. Individual stars can form along dense
filaments, usually in compact, dense cores within the filamentary
structure.

\begin{figure}
%\hskip 0.8in
\begin{center}
\includegraphics[scale=0.22]{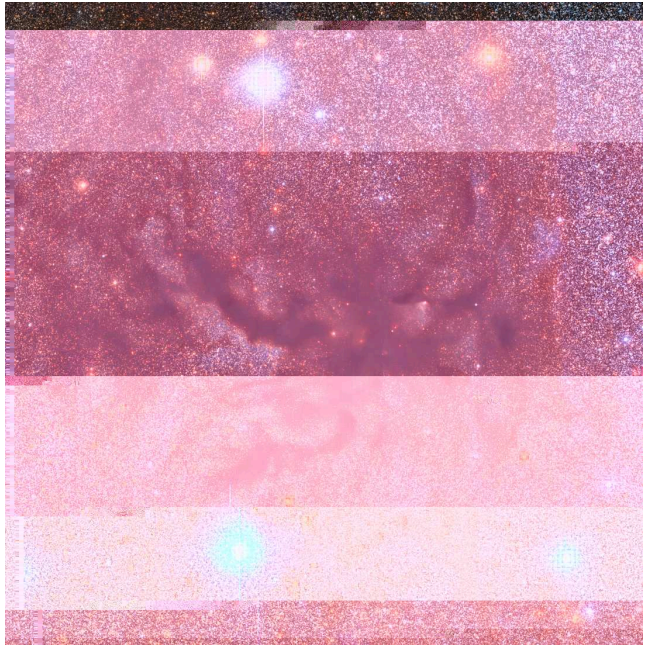}
\end{center}
\caption{Optical image of B59, a compact, dense core at the apparent
intersection of a number of filamentary structures and the site of the
formation of a small, embedded, optically invisible cluster. (Courtesy
of Alves, Vendame and Peris.) }
\end{figure} 

Clusters of stars typically form in the most massive and dense cores
within a GMC. Often, these massive cores are compact structures
located at the ends or at an intersection of filaments, as illustrated
in figure 5 (Myers 2009). The dense cores that form clusters range in
mass from $\sim 20$ to 2000 \msunsp and have sizes on the order of
1--2 pc in diameter (e.g., Higuchi et al. 2009). The spatial extents
of embedded clusters are comparable to the sizes of the dense cores
within which they reside. This suggests that the stars may be in
approximate virial equilibrium with the gas in the core. The cluster
members may thus not be colder dynamically than the surrounding
gas. On the other hand, there are some indications that the stars
within clusters may have started out dynamically colder than the
gas. Starless dense molecular cores, the precursors to individual new
stars, are characterized by subvirial velocities in a number of clouds
(e.g., Walsh \ea 2004; Peretto \ea 2006; Lada \ea 2008) and
protostellar objects have also been found to be spatially more
confined than more evolved young stellar objects (e.g., Teixeira \ea
2006; Muench \ea 2007). Such observations suggest that, at birth, new
stars in clusters are dynamically colder than the bulk of the gas
surrounding them. Such stars are then expected to violently relax and
collapse to smaller configurations. In doing so, dynamical
interactions can produce segregation of the most massive stars,
similar to observations (Allison \ea 2009). Comparison of the stellar
content of embedded clusters with the mass of their surrounding cores
yields star-formation efficiencies of typically between 10 and
30\%. If the cluster stars were indeed initially dynamical colder than
the bulk molecular gas, these measured `final' efficiencies will be
higher than the initial birth efficiencies (Lada \ea 1984; Goodwin \&
Bastian 2006). The observed low to modest final star-formation
efficiencies are key to understanding the early dynamical evolution
and infant mortality of such objects.

\begin{figure}
%\hskip 0.30in
\begin{center}
\includegraphics[scale=0.40]{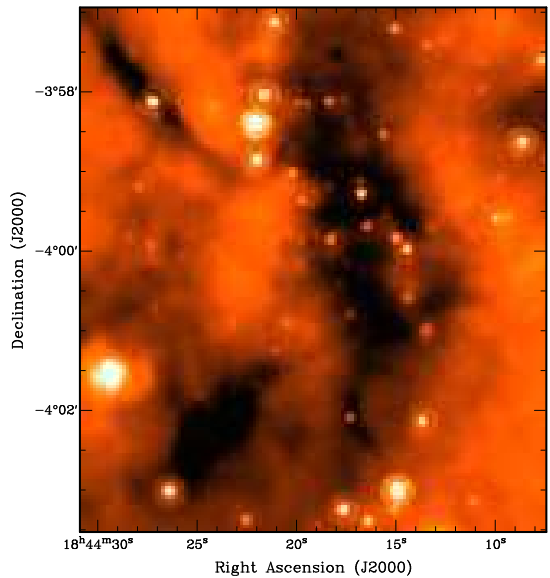}
\end{center}
\vskip -1.0in
\caption{Image of the infrared dark cloud G028.53--00.25, obtained at
24 $\mu$m with the {\sl Spitzer Space Telescope} (Rathborne \ea 2008).
This cloud contains in excess of 1000 \msunsp within a radius of a
parsec and yet exhibits no indication of active star formation. It is
possibly a future site of cluster formation. }
\end{figure} 

The most promising sites to investigate the formation of clusters may
be found in a class of objects known as infrared dark clouds (IRDCs)
(e.g., Rathborne \ea 2006). They are identified as high-extinction
dark clouds when viewed against bright, diffuse mid-infrared
(10--$20\mu$m) Galactic background emission. To be regions of high
extinction in the mid-infrared requires them to have enormous column
densities ($10^{23} - 10^{25}$ cm$^{-2}$) and correspondingly high
volume densities ($> 10^5$ cm$^{-3}$), just the type of conditions
that are ripe for star and cluster formation. The more massive of
these clouds tend to be concentrated in the so-called 5 kpc ring
(Simon \ea 2006), a large-scale Galactic structure rich in massive
molecular clouds and active star formation. Millimetre-wave studies
show many IRDCs to contain cores of sufficient mass (100--1000 \msun)
to form modest to rich clusters. Yet, many of these objects show no
evidence of significant star or cluster formation. Figure 6 shows one
of these IRDCs imaged with the {\sl Spitzer Space Telescope}. This
particular cloud contains more than $10^3$ \msun, but exhibits no
evidence for star formation. Its compact size and large mass suggest
that it is gravitationally bound and very likely a site of future
cluster formation. Rathborne \ea (2009) estimate that a 300 \msunsp
cluster could ultimately emerge from this cloud. Other massive IRDCs
have exhibited evidence for the formation of massive stars within
their boundaries. Since massive stars are rarely formed in isolation,
these clouds may also be the sites of incipient cluster formation
(Rathborne \ea 2006). Because these clouds are identified from
mid-infrared observations, which can only be obtained from space,
their existence has only been known for about a decade. Since the
massive IRDCs tend to be relatively distant, they are challenging to
observe and, as a consequence, little is known about their detailed
physical properties. Unfortunately, too little is known about the
detailed conditions in the clusterless IRDCs to provide any strong
constraints on the problem of cluster formation. However, future
observational capabilities provided by such facilities as the {\sl
Atacama Large Millimeter Array} and the {\sl James Webb Space
Telescope} hold the promise for obtaining a more detailed description
of the physical conditions in these objects and for deeper insights
into the cluster-formation problem.

\subsection{Internal structure}

Knowledge of the internal structure of an embedded cluster can provide
additional clues relating to its origin and early dynamical
history. For the youngest clusters, the spatial distribution of member
stars may be closely related to the primordial structure of the cold,
dense molecular gas from which the clusters formed. In their review of
embedded-cluster properties, Lada \& Lada (2003) identified two
structural types of clusters, those that exhibit extended, irregular
surface-density distributions, often with multiple peaks, and those
that appear to be relatively more compact and have well-defined
centrally condensed surface-density distributions with one dominant
peak. More recently, Ferreira \& Lada (2009, personal communication)
used ground-based infrared imaging observations to systematically
investigate the structure of approximately 40 previously catalogued
embedded clusters. They found that about 60\% of these objects were of
the centrally condensed variety. The remainder exhibit relatively flat
surface-density profiles with more irregular boundaries and more
elongated shapes than the centrally condensed clusters.

\begin{figure}
%\hskip 0.8in
\begin{center}
\includegraphics[scale=0.30]{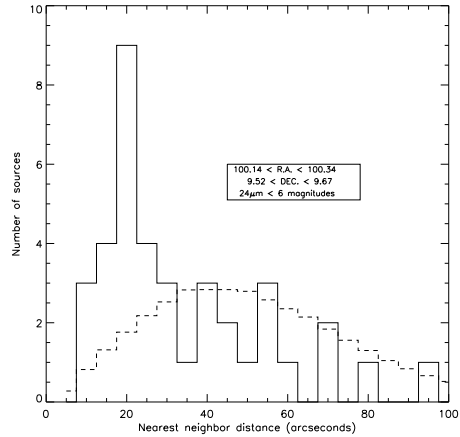}
\end{center}
\caption{Distribution of the nearest-neighbour separations of
protostars (solid trace) in an embedded cluster within NGC 2264,
compared to the expected distribution of nearest neighbours for a
random distribution of stars. The protostellar distribution is both
nonrandom and strongly peaked. The angular scale of this peak
corresponds to the expected Jeans length in the dense core containing
the cluster. (From Teixeira et al. 2006.)}
\end{figure} 

Observations with the {\sl Spitzer Space Telescope} at infrared
wavelengths of 3--24 $\mu$m provide an improved census of cluster
membership and thus enable better measurements of cluster shapes and
sizes. These observations have shown that most embedded clusters are
elongated, with a typical aspect ratio of about 2, very similar to the
aspect ratios of the filamentary molecular clouds in which the
clusters are born (Allen \ea 2007; Gutermuth \ea 2009). These
observations imply that the structures of most embedded clusters
probably reflect those of the original molecular clouds from which
they formed. They also indicate that understanding the origin of such
clusters is intimately tied to understanding the origin and evolution
of the dense molecular material that produces them.

A potentially significant clue concerning the nature of the physical
process of cluster formation is found in analysis of the distributions
of spatial separations of the youngest members of some embedded
clusters. Figure 7 shows the distributions of the nearest-neighbour
separations of protostars in an embedded cluster in the NGC 2264
star-forming complex (Teixeira \ea 2006). This distribution is
strongly peaked at an angular separation of 20 arcsec and clearly
different from that expected for a random distribution of
separations. The characteristic separation corresponds to the Jeans
length for the density and temperature of the cloud core containing
the cluster and suggests that Jeans fragmentation may be the physical
process that governs the formation of the cluster from the massive
cloud core. {\sl Spitzer Space Telescope} studies of other clusters
have found such evidence for Jeans fragmentation to be a more common
property of embedded clusters than previously thought (Gutermuth \ea
2009).

\section{Implications for understanding the origin and early evolution
of embedded clusters}

Because stars form in dense gas, it is the initial distribution of the
dense material in a cloud that controls the degree to which a cloud
forms an extended population of distributed stars and/or compact
clusters. That the mass spectrum of embedded clusters resembles the
shape of the mass spectrum of massive cores in GMCs appears to support
this surmise (Lada \& Lada 2003). Once dense gas is present, star
formation likely proceeds quickly. The timescale is somewhat uncertain
but the fact that numerous examples of massive, clusterless cores in
IRDCs exist probably indicates a time for the onset of cluster
formation on the order of a million years. On the basis of the ages of
stars in embedded clusters, the time needed to construct the full
cluster is likely 2--4 Myr from the onset of significant star
formation. The physical mechanism that causes a massive core to
fragment and form a cluster of stars is not clear. That the Jeans
scale appears to describe the separations of the young stellar objects
in embedded clusters suggests that a Jeans-like fragmentation process
is responsible for the formation of a stellar cluster from a massive,
dense core. Support for this idea also derives from measurements of
the initial mass function (IMF) of stars within individual clusters.
The stellar IMF has a characteristic mass of about 0.3 \msun, which
appears invariant among clusters and even stars in the field. This
suggests a mass scale for star formation that is consistent with
thermal Jeans fragmentation (Larson 1992). The Jeans instability
describes the situation where gravity overcomes the internal thermal
pressure of a gas cloud and leads to both its fragmentation and the
collapse of its individual fragments. However, it is not clear how
this mechanism would operate in a very massive core where internal
turbulence dominates thermal motions. Moreover, since the critical
conditions of density and temperature which determine the value of the
Jeans length and mass likely vary among regions, one would expect the
characteristic stellar mass of clusters to likewise vary, but it does
not.

Recent observations by the {\sl Chandra} and {\sl Spitzer Space
Telescopes} have revived the idea that external triggering could
represent another mode of cluster formation in the Galaxy. In the
Cepheus B (Getman \ea 2009) and W5 (Koenig \ea 2009) star-forming
complexes, spatiotemporal patterns of star formation have been found
in the distribution of recently formed stars. Such patterns of
sequential star formation have long been regarded as evidence for
triggered star formation (e.g., Elmegreen \& Lada 1977). Triggering
produces enhanced compression of existing molecular clouds, raising
their mean densities and, in doing so, increasing the fraction of
cloud mass above the critical densities necessary for star
formation. One way to compress a cloud is through increasing its
external pressure. It is possible that conditions in merging and
interacting galaxies, and in the nuclear regions of starburst
galaxies, produce globally increased external pressures that
significantly increase the fraction of dense material in a cloud and
lead to the formation of massive dense cores and clusters. Lack of
such conditions (i.e., high external pressures) may explain why very
massive clusters are not presently forming in the Milky Way.

Although it is beyond the scope of this particular review to discuss
the theory of cluster formation, it is useful to briefly mention the
results of recent calculations since some insight into the physical
process of cluster formation can be provided by numerical
simulations. A more detailed discussion of the theory of cluster
formation can be found in the accompanying review by Clarke
(2010). Simulations of cluster formation follow the collapse and
fragmentation of massive, initially turbulent clouds from some
specified initial gaseous state down to the formation of stellar
objects. The results of these calculations yield impressive details
concerning the clusters they form but are very sensitive to the input
physics. In particular, calculations starting with roughly similar
initial conditions (i.e., $\sim 50 - 100$ \msunsp turbulent clouds)
can have drastically different outcomes, depending on whether feedback
from newly formed stars is included. Models without feedback (e.g.,
Bate \& Bonnell 2005) produce highly filamentary cloud configurations
reminiscent of observed clouds. These models produce extensive
(efficient) fragmentation and form rich clusters of primarily low-mass
stars whose characteristic mass is directly related to the Jeans mass
and thus varies with initial conditions (i.e., density). Models that
include feedback (e.g., Krumholtz \ea 2007; Bate 2009) produce more
massive cloud cores, smaller clusters, more massive stars and an
invariant characteristic stellar mass, all in better agreement with
observations of embedded clusters. Most importantly perhaps, the
models which include feedback from the newly formed stars are also
characterized by low star-formation efficiencies, again similar to
observations.

Once embedded clusters are formed, the process of their subsequent
dynamical evolution and emergence from molecular clouds is relatively
well understood. Analytic and numerical calculations have shown that
the key factor in determining the evolution of an embedded cluster is
the (final) star-formation efficiency characterizing its formation
(e.g., Lada \ea 1984; Geyer \& Burkert 2001; Kroupa \& Boily
2002). Observations indicate that the star-formation efficiencies for
embedded clusters range between 10 and 30\% (Lada \& Lada 2003;
Higuchi \ea 2009) with some indications that those with the highest
efficiencies are the most evolved objects (Higuchi \ea
2009). Numerical simulations of cluster formation suggest that these
low efficiencies are caused by the radiative feedback of the recently
formed stars on their surrounding dense gas. In any case, the
gravitational glue that binds the stars and gas together in an
embedded cluster system is largely provided by the gas. We expect
that, as more and more stars form in the cluster, the feedback from
the stars due to radiation and energetic jets and outflows, generated
in the early stellar-accretion process, becomes destructive and begin
to disperse the dust and gas in the core. Massive stars are
particularly destructive and a single O star can completely disrupt an
entire cloud in a very short period of time. Because of the low
efficiency of conversion of gaseous to stellar mass in these systems,
they can easily be completely disrupted if the gas is removed on
timescales comparable to the dynamical time of the system. Thus, low
efficiency coupled with relatively rapid gas removal can explain why
90\% of embedded clusters do not survive their emergence from
molecular clouds as bound systems (Lada \& Lada 2003).

Paradoxically, the clusters that survive emergence from molecular
clouds as bound systems are likely the most massive systems at the
upper end of the embedded-cluster mass function. This is despite the
fact that these systems are the ones most likely to produce
destructive O stars. Simulations indicate that, as clusters are
disrupted by gas-removal processes, this disruption is rarely 100\%
complete and bound remnants containing a small fraction of the
original stellar mass are often left behind (Lada \ea 1984; Kroupa \&
Boily 2002). The size of the bound remnant depends on the initial size
of the cluster and on how rapidly gas removal occurs. Lada \& Lada
(2003) suggest that the field population of open clusters in the disc
of the Milky Way was likely provided by embedded clusters with masses
in excess of 500 \msun. This may also explain the relative lack of
truly massive (i.e., $M > 10^4$ \msun) clusters in the Milky Way with
ages in excess of 10 Myr. The few that are formed in the Milky Way do
not emerge completely unscathed. Only smaller remnants of these giants
survive as bound systems beyond 10 Myr.

\section{Concluding remarks}

More than two centuries after Herschel (1785) first considered the
question, we now understand that clusters are formed in cold, dark
molecular clouds. Stellar clusters begin their lives deeply buried in
dense molecular gas and dust as embedded infrared protoclusters. Over
the past two decades, advances in detector and telescope technology
have led to steady advances in our knowledge of embedded
protoclusters. These extremely young stellar systems appear to account
for a significant fraction of all star formation presently taking
place in the Milky Way and, thus, they are tracers of the current
epoch of star formation in the Galaxy. Although considerable progress
has been made towards understanding the basic physical properties of
embedded clusters, a physical theory of cluster formation eludes
us. In the Milky Way, the primary mode of cluster formation at the
present epoch appears to be one in which relatively compact, massive
dense cores in GMCs undergo a process of Jeans-like fragmentation that
transforms cold, dense interstellar material into stellar
form. Triggering of clouds by shocks or other mechanisms that lead to
an increase in external pressure may represent an additional mode of
Galactic cluster formation. Although its relative contribution to
present-day cluster formation in the Milky Way is unclear, this latter
mechanism may be significant in interacting galaxies and nuclear
starbursts, and perhaps even in the early history of the Milky Way
itself. It is now clear that to develop a predictive theory of cluster
formation will require both a better understanding of the process of
star formation and a comprehensive understanding of the physical
mechanisms that organize the dense material of a GMC into massive,
compact cores which, for a typical GMC in the Milky Way, occupy only a
small fraction ($< 1$\%) of its volume and account for only a small
fraction (1--10\%) of its total mass.

\begin{acknowledgements}
I thank Joana Ascenso, Ken Janes, Lisa Townsley, Eric Feigelson,
Elizabeth Lada and Bruno Ferreira for informative discussions, Joana
Ascenso for preparing figure 3 and Jo\~ao Alves for providing figure
5. I am grateful to Simon Goodwin and Richard de Grijs for
constructive comments that improved this review.
\end{acknowledgements}

\end{document}